\documentstyle[12pt]{article}
\begin{document}

\title{On the Absence of the Zeno Effect in Relativistic Quantum Field
Theory}
\author{Ram\'on F.
Alvarez-Estrada\thanks{E-mail:ralvarez@eucmax.sim.ucm.es} \\
 {\small\em Departamento
de F\'{\i}sica Te\'orica, Universidad Complutense, 28040 Madrid, Spain}
\and
Jos\'e L. S\'anchez-G\'omez\thanks{E-mail: jolu@delta.ft.uam.es} \\
{\small\em Departamento de F\'{\i}sica Te\'orica, Universidad Aut\'onoma
de Madrid, 28049-Madrid, Spain}}

\maketitle

 \begin{abstract}
We study the time evolution of decaying particles in
renormalizable models of Relativistic Quantum Field Theory.
Significant differences between the latter and Non Relativistic
Quantum Mechanics are found ---in particular, the Zeno effect seems to
be absent in such RQFT models. Conventional renormalization yields
finite time behaviour in some cases but fails to produce finite survival
probabilities in others.
\end{abstract}

The analysis of the finite time evolution in relativistic quantum field
theory (RQFT) and, in particular, the behaviour at short times of the
survival probability of unstable particles in RQFT is a relevant
matter as far as the foundations of quantum mechanics are concerned and
has some peculiarities which make it to be worth studying. But before
going to RQFT, we will briefly recapitulate the main features of the
survival amplitude (SA) at short times in non relativistic quantum
mechanics (QM henceforth); see, for instance, \cite{kn:fonda} for a
comprehensive review.

\bigskip
Let \(P(t)\,(t>0)\) the survival probability (SP) for an unstable system
represented in QM by some normalized state $\mid\!\Psi\rangle$, and let
H be the Hamiltonian governing the evolution of the quantum system. The
SP is given by
\(P(t) = \mid\langle\Psi\mid\exp(-itH)\mid\Psi\rangle\mid^{2}\)
($\hbar = 1$). Then, assuming the finiteness of the energy dispersion in
the initial (unstable) state, i.e.,
\begin{equation}
 (\Delta\!E)^{2} = \langle\Psi\!\mid\!H^{2}\!\mid\!\Psi\rangle -
(\langle\Psi\!\mid\!H\!\mid\!\Psi\rangle)^{2} < \infty,
\end{equation}
as indeed happens in QM, one can show that a very short times
\begin{equation}
P(t) \simeq 1 - (\Delta\!E)^{2}t^{2}\;\; (t\rightarrow 0^{+}).
\end{equation}
Notice that the last equation shows the violation of the exponential
decay law at very short times. The quadratic behaviour in $t$ gives rise
to the so-called quantum Zeno effect, whose consequences and possible
experimental detection \cite{kn:itano} have been profusely discussed
(see \cite{kn:pasca} and references therein for some recent work on this
effect).

\bigskip
Another interesting point --- which, however, is not the main subject
of this paper--- is the behaviour of the survival probability for large
t values. One knows that if the hamiltonian spectrum is bounded from
below, then the decay cannot be exponential, and
$P(t)>P_{exp}(t),\;\mbox{as }t\rightarrow\infty$.
Now, whereas that property (positivity of the
energy) always holds in QM, the situation is less clear in RQFT. Let
$H_{ren}$ be the hamiltonian of a certain model of RQFT, after having
carried out standard ultraviolet renormalization. To the best of our
knowledge, $H_{ren}$ has been {\em rigorously} constructed for
superrenormalizable models, like, for instance, the $\lambda\phi^{4}$ and
Yukawa RQFT, in 1+1 dimensions, but presumably not for renormalizable
models of the kind dealt with in this work. In the superrenormalizable
RQFT models for which $H_{ren}$ exists, the latter is lower bounded
\cite{kn:glimm}. In any case, it is not the lower boundedness of the
hamiltonian but, rather, the finiteness of the energy dispersion in the
initial (unstable) state what is relevant for the existence of the Zeno
effect.

\bigskip
Regarding the RQFT case, it has been pointed out \cite{kn:maiani}
that the unrenormalized survival amplitude of an unstable particle has,
in any model of RQFT (superrenormalizable or renormalizable), a singular
behaviour as $t\rightarrow 0^{+}$ and, in order to provide a cure for
it, a new characteristic time has been introduced. Such singularities
seem to be related to those
already discussed by Bogoliubov and Shirkov \cite{kn:bogo} for the
Schr\"{o}dinger state
vector in RQFT. More recently, the present authors have shown that in a
superrenormalizable model of RQFT (namely, a direct extension of the
scalar $\phi^{3}$-theory to the decay of an unstable particle) the above
short-time singularities of the survival amplitude can be fully
eliminated through a conventional renormalization procedure
\cite{kn:AESG}. Here, we are going to analyze the time evolution of the
SP of unstable particles in renormalizable models of RQFT. Now, one
should be
careful about vacuum polarizing interactions and ultraviolet divergent
renormalization. Let us clearly state the problem

\bigskip
We consider ---with respect to a given reference frame--- some ``free"
Hamiltonian $H_{0}$ which describes particles for $t\leq 0$ in RQFT in
the Schr\"{o}dinger picture. Also, let $\vec{P}$ the total three-momentum
operator, with $[\vec{P}, H_{0}] = 0$, and let
$\mid\!i,\vec{p_{i}}\rangle$
the initial state representing the unstable particle at $t = 0$, with
four-momentum $p_{i} = (E_{i},\vec{p_{i}})$, $E_{i} = (m_{i}^{2} +
\vec{p_{i}}^{2})^{1/2}$, and renormalized mass $m_{i}$. Thus,
$\mid\!i,\vec{p_{i}}\rangle$ is a common eigenstate of both $H_{0}$ and
$\vec{P}$.
If $\mid\!0\rangle$ represents the vacuum of $H_{0}$, then
\begin{equation}
\mid\!i,\vec{p_{i}}\rangle \equiv \mid\!i\rangle = a^{+}(\vec{p_{i}})
\mid\!0\rangle ,
\end{equation}
where $a^{+}(\vec{p_{i}})$ is the associated creation operator (in the
Schr\"{o}dinger picture).
The interaction Hamiltonian $H_{I}$, which acts for $t>0$ and is
responsible for the decay has, by assumption, the following properties:

\bigskip
\noindent 1. It commutes with $\vec{P}$.

\bigskip
\noindent 2. It gives rise to a renormalizable local RQFT.

 \bigskip
\noindent 3. It involves a small coupling constant, in the sense that
one can reliably apply perturbation theory.

\bigskip
Upon going over to the interaction picture (ip) and after some
straightforward calculations, we get the basic formula enabling us to
compute the survival amplitude (SA) in renormalized perturbation theory
\begin{eqnarray}
A(t)&=& \langle i\!\mid\!i\rangle^{-1}
\sum_{n=0}^{\infty}\frac{(-i)^{n}}{n!}. \nonumber  \\
& & \int_{0}^{t}dt_{1}\cdots\int_{0}^
 {t}dt_{n}\:\langle 0\!\mid T[a(t)H_{I}(t_{1})\ldots
H_{I}(t_{n})a^{+}(0)]\mid\!0\rangle
\end{eqnarray}
where, for simplicity, we have omitted ip in the operators, as well as
$\vec{p_{i}}$ in $a$ and $a^{+}$. Of course, T stands for time-ordered
product. One can now evaluate $A(t)$ by using (4) and Wick's theorem in
the standard way. Then it will be useful to introduce the following
factorized form of the SA
\begin{equation}
 A(t) = \exp(-iE_{i}t)\tilde{A}(t),
\end{equation}
where the ``reduced" survival amplitude $\tilde{A}(t)\rightarrow 1$ for
any $t > 0 \;\mbox{if }H_{I}\rightarrow 0$.
 Now we have to deal with the divergences generically known as
disconnected vacuum contributions (dvc). For the sake of brevity, we do
not present here the details of the calculation (see \cite{kn:AESG} for
a more elaborated treatment), which yields the following result
\begin{equation}
\tilde{A}(t) = A_{dvc}(t)A_{ph}(t).
\end{equation}
$A_{ph}(t)$, to be called hereafter the physical survival amplitude, is
the sum of all perturbative contributions to $\tilde{A}(t)$ in each of
which $a(t)$ is not contracted with $a^{+}(t = 0)$. Therefore all the
contributions to $A_{ph}$ are free of infinite volume divergences; all
the latter are included in the factor $A_{dvc}(t)$. We should notice
that $A_{ph}\rightarrow 1 \;\mbox{as }H_{I}\rightarrow 0$. We shall
concentrate on $A_{ph}$. As already said, the above procedure has been
applied to the case of a superrenormalizable ($\phi^{3}$-like)
interaction \cite{kn:AESG}, having shown that, after a standard
ultraviolet mass renormalization, the survival amplitude is finite for
all
$t$ and it behaves linearly in $t$ as $t\rightarrow 0$ (no Zeno effect,
then). Nevertheless, the scope of such superrenormalizable interactions
is rather limited, so that, here, we are going to deal with the case of
renormalizable interactions, which is more relevant from a physical
point of view.

\bigskip
    We shall consider a (unstable) relativistic spin 1/2
particle (i), with renormalized mass $m_{i}$, which can decay into two
relativistic particles $(a,\,b)$, with renormalized masses
$m_{a},\,m_{b}$, so that: i) $m_{i}>m_{a}+m_{b}$, and ii) $b$ is a
spin-1/2 fermion while $a$ is a scalar spinless boson (we are
considering all the particles neutral for simplicity; the general case
is physically equivalent).
Now $H_{0}$ is the
sum of three free Hamiltonians: one, corresponding to particle $a$ being
similar to those in the superrenormalizable case (kinetic energy of a
scalar spin-0 boson) and the other two, of particles $i$ and $b$, being
the well-known for relativistic free Dirac fermions.  For $t>0$ the
interaction Hamiltonian in the Schr\"{o}dinger picture is
\begin{eqnarray}
H_{I} & = & \lambda\int
d^{3}\vec{x}\:[\bar{\Psi}_{b}(x)\Psi_{i}(x)\Phi_{a}(x)] +\mbox{h.c.} -
 \delta m_{i}\int d^{3}\vec{x}\:N[\bar{\Psi}_{i}(x)\Psi_{i}(x)]-
\nonumber \\
      &    & (Z_{i,2}-1)\int d^{3}\vec{x}\:[\bar{\Psi}_{i}
(\gamma.\partial)\Psi_{i}- m_{i}\bar{\Psi}_{i}\Psi_{i}] +
\mbox{o.c.t.}
\end{eqnarray}
where $\lambda$ is a renormalized dimensionless coupling constant. The
mass renormalization counterterm for the unstable i-fermion is, to
order $\lambda^{2}$
\begin{equation}
\delta m_{i} = \frac{i\lambda^{2}}{(2\pi)^{4}}\int
d^{4}k\:\frac{-k.\gamma+m_{b}}
{(k^{2}-m_{a}^{2}+i\epsilon)(k^{2}-m_{b}^{2}+i\epsilon)}
\end{equation}
The term proportional to $Z_{i,2} - 1$ is the standard contribution
corresponding to the wave-function renormalization for particle $i$,
and o.c.t. denote other renormalization counterterms which are not
relevant here. The calculation now turns out to be much more elaborated
than in the superrenormalizable case. One has, to order $\lambda^{2}$,
\begin{equation}
A_{ph}(t) - 1 = \frac{m_{i}}{E_{i}}\:
\bar{u}(\vec{p_{i}},\sigma)K_{ren}u(\vec{p_{i}},\sigma)
\end{equation}
 \begin{equation}
K_{ren} = K + i\delta m_{i}t+i(Z_{i,2}-1)(\gamma.p_{i}-m_{i})t
\end{equation}
\begin{eqnarray}
K & = & \frac{\lambda^{2}}{(2\pi)^{5}}\int d^{4}k_{1}d^{4}k_{2}\;
\frac{\gamma.k_{2}+m_{b}}{(k_{1}^{2}-m_{a}^{2}+i\epsilon)(k_{2}^{2}-
m_{b}^{2}+i\epsilon)}\delta^{(3)}(\vec{k_{1}}+\vec{k_{2}}-
\vec{p_{i}}).  \nonumber \\
  &   & \frac{\exp[it(E_{i}-k_{1}^{0}-k_{2}^{0})]-1}{i(E_{i}-k_{1}^{0}-
k_{2}^{0})}.\:\frac{\exp[it(k_{1}^{0}+k_{2}^{0}-E_{i})]-1}{i(k_{1}^{0}+
k_{2}^{0}-E_{i})},
\end{eqnarray}
where $\epsilon\rightarrow 0^{+}$ and $u,\;\bar{u}$ denote
normalized Dirac spinors for the unstable particle with polarization
$\sigma$.
In K, we integrate over $\vec{k_{2}}$ and, by adding
$-i\epsilon$ to $E_{i}$
(which is a consistent procedure), we perform a residue integration over
$k_{2}^{0}$. These integrations give K as a piece linear in t plus
another contribution, $\Xi$. In $\Xi$, we integrate over $k_{1}^{0}$ by
residues, then obtaining
\begin{equation}
K_{ren} = it\Sigma_{ren} + \lambda^{2}\Xi
\end{equation}
\begin{eqnarray}
\Sigma_{ren}-[\delta m_{i} + (\gamma.p_{i}-m_{i})(Z_{i,2}-1)] =
\nonumber \\
-i\frac{\lambda^{2}}{(2\pi)^{4}}\int\frac{d^{4}k_{1}}{k_{1}^{2}-m_{a}^{2}
+i\epsilon}.\:\frac{\gamma.(p_{i}-k_{1})+m_{b}}{(p_{i}-k_{1})^{2}-m_{b}^2+
i\epsilon}\equiv \Delta + iAbs\Sigma_{ren}
\end{eqnarray}
\begin{eqnarray}
\Xi = \int
\frac{d^{3}\vec{k_{1}}}{(2\pi)^{3}}\frac{1}{4E_{a}E_{b}}\{\frac{[\exp
it(E_{i}-E_{a}-E_{b})-1][E_{b}\gamma^{0}+(\vec{k_{1}}-\vec{p_{i}}).\vec
{\gamma}+m_{b}]}{(E_{a}+E_{b}-E_{i}-i\epsilon)^{2}} + \nonumber \\
 \frac{[\exp
it(E_{i}+E_{a}+E_{b})-1][-E_{b}\gamma^{0}+(\vec{k_{1}}-\vec{p_{i}}).
\vec{\gamma}+m_{b}]}{(E_{a}+E_{b}+E_{i}-i\epsilon)^{2}}\}
\end{eqnarray}
where \(E_{a} =
(m_{a}^{2}+\vec{k_{1}}^{2})^{1/2},\;E_{b}=[m_{b}^{2}+(\vec{p_{i}}-
\vec{k_{1}})^{2}]^{1/2}\).

\bigskip
$\Xi$ is the sum of two contributions each of which seems
logarithmically ultraviolet divergent through naive power counting.
Fortunately, $\Xi$ is finite for any $t\geq 0$, since each divergence
proportional to $\vec{\gamma}$ cancels out by symmetric integration,
while those proportional to $\gamma^{0}$ cancel out with one another.
$\Sigma_{ren}$ is the renormalized self-energy of the unstable particle.
The quantities $\Delta$ and $Abs\Sigma_{ren}$ (the absorptive part of
$\Sigma_{ren}$), defined uniquely through standard methods, are finite.
We shall define the quantity $\Xi_{1}$ out of the first term on the
right hand side of Eq. (14) as follows: i)we replace $\exp
it(E_{i}-E_{a}-E_{b})-1$ by $i\sin t(E_{i}-E_{a}-E_{b})$, ii) if
$A\equiv E_{a}+E_{b}-E_{i}$, we replace $(A-i\epsilon)^{-2}$ by
$-i\pi\delta'(A)$. (Recall that $(A-i\epsilon)^{-2} =
PA^{-2}-i\pi\delta'(A)$, where P is the principal part and $\delta'$
stands for the derivative of the $\delta$ function.)This gives $\Xi =
\Xi_{1}+\Xi_{2}$, where $\Xi_{2}$ embodies all the remaining
contributions to $\Xi$.
Some direct computations yield the following properties:

\bigskip
\noindent 1) After cancelling out a linearly ultraviolet divergent piece
proportional to $\vec{\gamma}$ (by symmetric integration), $\Delta$
bears the form $\Delta_{1}+\gamma^{\mu}\Delta_{2,\mu}$, $\Delta_{1}\;
\mbox{and }\Delta_{2,\mu}$ being logarithmically divergent quantities.

\bigskip
\noindent 2) By employing $\delta m_{i}$ and the corresponding
logarithmically ultraviolet divergent expression for $Z_{i,2}-1$ to
order $\lambda^{2}$ (see, for instance, \cite{kn:iz}), all
logarithmically ultraviolet divergences cancel out in \( X\equiv\delta
m_{i}+(\gamma.p_{i}-m_{i})(Z_{i,2}-1)+\Delta\), which becomes finite.
Moreover $\bar{u}(\vec{p_{i}},\sigma)X u(\vec{p_{i}},\sigma)$ is real.

\bigskip
\noindent 3) $\Xi_{1}$ is linear and t, and, moreover, \(-t
Abs\Sigma_{ren}+ \lambda^{2}\Xi_{1}=0\)

\bigskip
\noindent 4) $\Xi_{2}$ is ultraviolet finite

\bigskip
We shall define $\Xi_{3}$ out of $\Xi_{2}$ as follows: in the
expression yielding the latter, we replace \(\exp
it[E_{i}\mp(E_{a}+E_{b})]-1\) by \(\cos t[E_{i}\mp(E_{a}+E_{b})]-1\).
Then  we have $\Xi_{2}=\Xi_{3}+\Xi_{4}$, where
\begin{eqnarray}
\Xi_{3}&=& \frac{1}{(2\pi)^{3}}\int\frac{d^{3}\vec{k_{1}}}{4E_{a}E_{b}}
\{\frac{\cos
t(E_{i}-E_{a}-E_{b})-1}{(E_{a}+E_{b}-E_{i})^{2}}\:[E_{b}\gamma^{0}+(\vec{k_
{1}}-\vec{p_{i}}).\vec{\gamma}+m_{b}] +  \nonumber \\
& &\frac{\cos t(E_{i}+E_{a}+E_{b}) -1}{(E_{a}+E_{b}+E_{i})^{2}}\:[-E_{b}
\gamma^{0}+(\vec{k_{1}}-\vec{p_{i}}).\vec{\gamma}+m_{b}]\}
\end{eqnarray}
while $\Xi_{4}$ contains all the remaining contributions to $\Xi_{2}$.
Then
\[K_{ren}= itX + \lambda^{2}\Xi_{4} +\lambda^{2}\Xi_{3}\]

The key reasons for the successive definitions of \(\Delta,\;\Xi_{j},\;
j=1,2,3,4\) are the above results 1)-- 4) plus the following ones:

\bigskip
\noindent 5) \( \bar{u}(\vec{p_{i}},\sigma)[itX + \lambda^{2}\Xi_{4}]
u(\vec{p_{i}},\sigma)\) is purely imaginary,

\bigskip
\noindent 6) \( \bar{u}(\vec{p_{i}},\sigma)\lambda^{2}\Xi_{3}u(\vec{p_{i}},
\sigma)\) is real

\bigskip
  Consequently, the physical survival probability up to order
$\lambda^{2}$ is
\begin{equation}
 P(t) = \mid A_{ph}(t)\mid^{2} = 1 +
2\lambda^{2}[\bar{u}(\vec{p_{i},\sigma})\Xi_{3}u(\vec{p_{i}},\sigma)]\:
\frac {m_{i}}{E_{i}}.
\end{equation}

We shall now deal with a very important physical requirement, namely,
the consistency between (16) and Fermi's golden rule. For that purpose
we let $t\rightarrow\infty$ and recall that, then,
\(q^{-2}[1 - \cos(tq)]\rightarrow \pi t\delta(q)\) for any real $q$. A
glance at Eq. (15) shows that only the first fraction on its r.h.s.
yields a non-vanishing contribution as $t\rightarrow\infty$. Now, using
well-known properties of Dirac matrices and spinors, one readily
gets
\begin{equation}
P(t) \simeq 1-\frac{m_{i}}{E_{i}}\:\Gamma t\;\;
(t\rightarrow\infty),
\end{equation}
where $\Gamma$ can be writen as
\begin{equation}
\Gamma =\frac{
\lambda^{2}}{16\pi^{2}m_{i}}\int\frac{d^{3}\vec{k_{1}}d^{3}
\vec{k_{2}}}{E_{a}(\vec{k_{1}})E_{b}(\vec{k_{2}})}(p_{i}k_{2}+m_{i}m_{b})
\delta^{4}(p_{i}-k_{1}-k_{2}).
\end{equation}
Notice that $\Gamma$ is relativistically invariant and coincides exactly
(as a straightforward computation employing Dirac matrices and their
traces show) with the total decay rate of the unstable particle in its
rest frame ($\vec{p_{i}}=0$).
By taking into account that $m_{i}/E_{i} = (1-\vec{v_{i}}^{2})^{1/2}$,
$\vec{v_{i}}$ being the velocity of the decaying particle, and
introducing its lifetime $\tau=\Gamma^{-1}$ at rest, Eq. (17) can also
be
cast as \(P(t) = 1- t(1-\vec{v_{i}}^{2})^{1/2}/\tau\), which is simply
showing the relativistic fact that an unstable particle with velocity
$\vec{v_{i}}$ in an inertial frame has a lifetime
$\tau(1-\vec{v_{i}}^{2})^{-1/2}$ in that frame. Then, Eqs. (17-18) are
consistent with the linearization of the exponential decay behaviour
$\exp[-t(1-\vec{v_{i}}^{2})^{1/2}\Gamma]$ in an inertial frame;
$\Gamma$,
 the decay rate in the rest frame, being given by Fermi's golden rule

 \bigskip
Next, we shall investigate the behaviour of the survival probability in
the rest frame ($\vec{p_{i}}=0$) for small positive values of $t$. For
that purpose, we write $\Xi_{3}= Y_{+}+\gamma^{0}Y_{-}$ (noticing
that the contribution proportional to $\vec{\gamma}$ vanishes by
symmetric integration). We are interested in the behaviours of $Y_{\pm}$
for small positive $t$. Such behaviours are determined by large values
of $\mid\!\vec{k_{1}}\!\mid$. It suffices to neglect all masses in
$Y_{+}$, and one readily obtains
\( Y_{+}\sim  -m_{b}t/(16\pi)\).
The treatment of $Y_{-}$ requires more care, due to an additional factor
$E_{b}(\vec{k_{1}})$ which appears with opposite signs in both
contributions in the r.h.s. of (15) and, hence, leads to cancellations
at
large $\mid\!\vec{k_{1}}\!\mid$. One finds \(Y_{-}\sim -tm_{i}/(32\pi)\).
Then, for very small values of $t$, the survival probability is given by
\begin{equation}
 P(t) \simeq 1- \frac{\lambda^{2}}{16\pi}\:(2m_{b}+m_{i})t
\end{equation}
The linear behaviour for very small values of $t$ appearing in (19)
shows that the Zeno effect is not present.

\bigskip
We shall now consider another renormalizable model, similar regarding
physical description and notations to the one previously described,
except for the following crucial differences:

\bigskip
\noindent a) The unstable particle  $i$ is a spinless boson
which decays into two spin 1/2 (Dirac) fermions $(a,\:b)$ (again, all
particles are neutral for simplicity).

\bigskip
\noindent b) For $t>0$, the interaction hamiltonian is
\begin{equation}
H_{I} =
\lambda\int d^{3}(\vec{x})\:[\bar{\Psi}_{a}(\vec{x})\Psi_{b}(\vec{x})
\Phi_{i}(\vec{x})]\,+\:\mbox{h.c.}\,+\:\mbox{c.t.}
\end{equation}
The counterterms (c.t.) include both the mass and wave function
renormalization counterterms for the decaying boson (see, for instance,
\cite{kn:lurie} for their detailed expressions) plus another
renormalization counterterms which are not relevant for our calculation.
We shall limit ourselves to study the physical survival amplitude
$A_{ph}(t)$ for the decaying boson in its rest frame ($\vec{p_{i}}=0$).
A computation entirely similar to the one carried out for the previous
model yields
\begin{equation}
A_{ph}(t) \simeq 1+ \frac{1}{2m_{i}}(iXt+\lambda^{2}\Xi_{4}+\lambda^{2}
\Xi_{3}).
\end{equation}
The quantities X and $\Xi_{4}$ are the respective analogues of those for
the previous model: both $itX$ and $\Xi_{4}$ are pure imaginary. X is
the ultraviolet finite contribution which results after cancellations
between: i) divergent terms arising from the actual analogue of K (Eq.
(11)
for the previous model) and the mass and wavefunction
renormalization counterterms for
the decaying boson, ii) finite contributions linear in $t$ (analogues of
$-t\:Abs\Sigma_{ren}$ and $\lambda^{2}\Xi_{1}$ in the previous model).
We shall concentrate in the relevant piece, which is real and, hence,
contributes to the survival probability
$P(t) =\:\mid\!A_{ph}(t)\!\mid^{2}$:
\begin{eqnarray}
\Xi_{3} &=&
\frac{1}{(2\pi)^{3}}
\int\frac{d^{3}\vec{k_{1}}}{4E_{a}
E_{b}}(m_{a}m_{b}+E_{a}E_{b}+\vec{k_{1}}^{2}). \nonumber \\
& & [\frac{\cos t(E_{a}+E_{b}-m_{i})-1}{(E_{a}+E_{b}-m_{i})^{2}}\:+\:
 \frac{\cos t(E_{a}+E_{b}+m_{i})-1}{(E_{a}+E_{b}+m_{i})^{2}}],
\end{eqnarray}
where $E_{a,b} = (m_{a,b}^{2}+\vec{k_{1}}^{2})^{1/2}$.

\bigskip
The crucial difference with respect to the previous model is that, now,
$\Xi_{3}$ is ultraviolet divergent for any finite value of $t>0$, as a
direct power counting shows. The same statement holds, in principle, for
$\Xi_{4}$. To the best of our knowledge, and in spite of the fact that
the present model is renormalizable, no cure seems to exist for such
divergences. It is not hard to see that the source of the difficulties
is related to the fact that the self-energy diagram for the boson is
quadratically ultraviolet divergent. Therefore, it seems that if the
relevant self-energy diagrams are, at most, logarithmically divergent,
then one can obtain a finite survival probability for any $t\geq 0$.
Contrarily, if higher order divergences are present, the survival
probability does not exist, in the present context of strictly RQFT, for
any finite value of $t>0$, and, then, a different
regularization procedure, such as the one introduced in \cite{kn:maiani}
---which, in a strict sense, lies outside the province of RQFT---should
be employed
in order to define the physical survival probability for finite times.
In any case, one should note that it is formally possible to let
$t\rightarrow\infty$ in (22), and employ
$[1-\cos(tq)]/q^{2}\:\rightarrow
\pi t\delta(q)$, as in the previous case. This formal limiting process
in the divergent (real) quantity $\Xi_{3}$ leads to a well defined
limit for the survival probability $P(t)$, namely, the actual analogue
of Eq. (17) holds as well in the present model: the actual $\Gamma$
coincides with the total decay rate of the unstable boson in its rest
frame.

\bigskip
Going back to the question of quadratic divergences and the
non-existence then ---in a strict RQFT context---of the survival
probability at finite times, it is worth recalling that, in QED, the
photon self-energy (vacuum polarization) diagram is quadratically
divergent by naive power counting but, actually, it turns out to be only
logarihmically divergent (before renormalization)
because of gauge invariance (see \cite{kn:iz}, for instance) ---a
feature which bears some relationship to the behaviour of the $W^{\pm}\;
\mbox{and } Z^{0}$ self-energies in the Weinberg-Salam model. Therefore
one could speculate on whether gauge invariance should play a crucial
role to properly analyze time evolution, and in particular the survival
probability at finite times of unstable particles in a RQFT context.

\bigskip
As a final comment, we would like to emphasize that our analysis has
relied on computations carried out in renormalized perturbation theory.
On the other hand, several renormalizable models on RQFT (like the
$\lambda\phi^{4}$ one) are accepted to have a trivial continuum limit
\cite{kn:mont} in a non-perturbative (renormalization group) framework.
In the latter case, the peculiarities of survival amplitudes of unstable
particles at finite times, as discussed in this work, would become
irrelevant. Here again, (non-abelian) gauge invariance might come to the
rescue, at least for QCD: it does make the renormalized theory non
trivial in the continuum limit and one could speculate that, upon
forcing all ultraviolet divergences to be logaritmic, time
evolution would be well defined.

\bigskip
\bigskip
One of us (RFAE) acknowledges the financial support of CICYT (Proyecto
AEN97-1693), Spain, and Human Capital and Mobility Programme, European
Comission (Contract  ERB CHRXCT 940423) Brussels; the other wishes to
thank CICYT (Pr. PB94-0194) for the same reason. We are indebted
to A. Dobado, V. Mart\'{\i}n Mayol, C. P\'erez
Mart\'{\i}n, F. Ruiz Ruiz and F.J. Yndur\'ain for discussions and useful
information.

     \end{document}